\documentclass[aps,pre,twocolumn,superscriptaddress,showpacs,showkeys ]{revtex4-2}
\usepackage{color}
\usepackage{graphicx}
\usepackage{hyperref}
\usepackage{amsmath,amssymb}
\usepackage{epstopdf}
\usepackage{subcaption}
\graphicspath{{figs/}}
\usepackage{caption}
\usepackage{dcolumn}   % needed for some tables
\usepackage{bm}        % for math
\usepackage{verbatim}   % for math

\begin{document}
\title{Fractal geometry-governed oxygen diffusion: Tumors vs. Normal 
Tissues}

\author{N. Valizadeh}
\email{valizadeh.neda1204@gmail.com}
\affiliation{Department of Physics, University of Mohaghegh Ardabili, P.O. Box 179, Ardabil, Iran}

\author{R. Rahimi}
\email{rrahimi@som.umaryland.edu}
\affiliation{University of Maryland School of Medicine, MD, United States of America}

\author{Ramin Abolfath}
\email{ramin1.abolfath@gmail.com}
\affiliation{Department of Physics and Astronomy, Howard University, Washington, DC 20059, United States of America}

\date{\today}

\begin{abstract}
{\bf Purpose}:
To develop a geometry-governed diffusion framework that explains differential tissue response under FLASH ultra-high dose rate (UHDR) irradiation by explicitly accounting for structural heterogeneity and anomalous transport in biological tissues.

{\bf Methods}:
We formulate a generalized diffusion--reaction model on fractal substrates to describe molecular transport in heterogeneous media. Tissue architecture is characterized by a fractal (Hausdorff) dimension \(D\), while scale-dependent transport inefficiency and memory effects are captured by a fractional parameter \(\theta\). Analytical solutions for radially symmetric geometries are derived and compared with classical normal (Euclidean) diffusion and a Gaussian reference model under identical physical conditions. Transport behavior is quantified through transient probability distributions and steady-state spatial profiles.

{\bf Results}:
The model reveals systematic suppression of long-range transport and enhanced localization as tissue structural complexity increases. Increasing \(\theta\) leads to subdiffusive dynamics, reduced effective diffusion lengths, and persistent non-Gaussian concentration profiles, even in the steady state. While increasing \(D\) alone enhances spatial accessibility, fractional dynamics dominate transport behavior when \(\theta>0\), counteracting geometric connectivity. These effects produce a separation between regimes characterized by efficient inter-track overlap and rapid homogenization, and regimes marked by isolated, long-lived reactive domains.

{\bf Conclusions}:
Fractal geometry provides a unifying physical framework for understanding tissue-dependent transport and differential response under FLASH UHDR irradiation. Normal tissues, characterized by near-Euclidean geometry and weak anomalous effects, permit greater inter-track interaction and recombination, whereas tumor-like tissues with elevated structural complexity exhibit localized transport and reduced collective chemical reactivity. This proof-of-principle study establishes tissue architecture as a fundamental determinant of transport efficiency and offers a mechanistic basis for experimentally observed FLASH tissue sparing, motivating geometry-aware modeling of radiobiological response.
\end{abstract}
%\pacs{05., 05.20.-y, 05.10.Ln, 05.45.Df}

\maketitle
\section{Introduction}

Diffusion in biological systems is rarely simple or uniform. Classical Fickian diffusion, formulated for homogeneous Euclidean continua, assumes a linear growth of mean-square displacement of ions with time, $\langle r^2(t) \rangle \propto t$, reflecting homogeneous diffusivity and/or ionic conductivity that relies on translational and rotational invariance of the Brownian motion in the medium. Yet, most natural and biological media are neither homogeneous nor Euclidean; rather, they are structurally complex, hierarchically organized, and dynamically heterogeneous, often exhibiting self-similar (fractal) features across spatial scales. In such environments, diffusion deviates from uniform behavior and becomes \textit{anomalous}---typically subdiffusive---following power-law scaling of the form
\begin{equation}
\langle r^2(t) \rangle \propto t^{\frac{2}{2+\theta}}, \quad \text{with } \theta > 0,
\label{eq1}
\end{equation}
where the exponent $\theta$ encodes the degree of geometric constraint, connectivity loss, and structural disorder of the medium. These anomalous transport patterns arise when the underlying structure itself possesses fractal geometry---characterized by a non-integer Hausdorff (fractal) dimension $D$ and associated spectral or walk dimensions that govern scaling of transport processes \citep{Mandelbrot1967,Feder1988,Bunde1991,Metzler2000}.
With a constant $\theta$ (including $\theta=0$), a decrease in $D$ from the Euclidean dimension $d$ effectively reduces the global transport efficiency, reflecting increased geometric constraints and reduced connectivity of diffusion pathways.
In these settings, diffusion is intrinsically non-Fickian: particle spreading depends not on a constant transport coefficient but on the topology, tortuosity, and percolation properties of the medium \citep{Tarasov2006,Zaslavsky2002}.

Biological tissues constitute archetypal fractal media. Cellular organization, extracellular matrix (ECM), and vasculature together form anisotropic, disordered, and multiscale networks whose geometry critically shapes transport phenomena \citep{Weibel1962,Baish2000,Gazit1995}. Tumor tissues, in particular, show a prominent architectural irregularity. Its microvasculature is chaotically branched and spatially heterogeneous, exhibiting fractal dimensions in the range $D \approx 2.1$--$2.8$, depending on the imaging modality and scale \citep{Gazit1997,Guiot2006}. This irregularity compromises both perfusion and diffusion, producing patchy oxygenation patterns and persistent hypoxic niches that cannot be captured by classical continuum diffusion models \citep{Grimes2014,Vaupel2021}. High-resolution oxygen-tension mapping further reveals sharp spatial fluctuations in $\mathrm{pO_2}$ distributions, with hypoxic and normoxic regions coexisting within micrometer distances \citep{Vaupel2021,Vaupel2004}. As tumor size increases, oxygen gradients flatten while remaining uniformly low, indicating chronic diffusion-limited hypoxia in contrast to the Krogh-type continuum transport assumptions \citep{Grimes2014}. The fractal dimension of the vascular or oxygen maps therefore provides a quantitative measure of the inefficiency of tissue space-filling and serves as a predictor of severity of hypoxia and resistance to treatment \citep{Degner1988,Baish2000,Lagerlöf2014}.

Fractal geometry likewise governs molecular transport and drug delivery within tumors. The ECM constitutes a disordered porous network that induces anomalous—often subdiffusive—propagation of therapeutic agents \citep{Netti1995,Stylianopoulos2013}. Computational and experimental studies demonstrate that densely yet chaotically vascularized tumors rapidly drain injected drugs, leading to poor intratumoral retention and spatially heterogeneous exposure \citep{Mohammadi2023,Kashkooli2021}. Strategies such as transient vessel normalization effectively reduce the geometric complexity of perfusion networks, thereby improving the uniformity of drug penetration and therapeutic efficacy \citep{Jain2005,Dewhirst2017}. Thus, geometry governs a fundamental determinant of transport efficiency, linking multiscale structural disorder to pharmacokinetic failure.

Beyond vascular and extracellular transport, experimental evidence indicates that membrane-scale heterogeneity further modulates diffusion-controlled processes in cancer. Viscosity-sensitive fluorescent probes and nanoparticle-based sensors reveal that tumor cell membranes exhibit altered lipid packing, spatially heterogeneous nanoviscosity, and markedly different viscoelastic properties relative to normal cells \citep{Li2023,Ober2019}. These membrane-level changes enhance lateral diffusion of lipids and reactive intermediates within the membrane plane and modify reaction–diffusion coupling at subcellular scales, directly impacting radical recombination kinetics under irradiation \citep{Vozenin2024,Favaudon2025}. Complementary biochemical analyses further demonstrate that cancer cells actively reprogram lipid composition—enriching polyunsaturated fatty acids and altering saturation profiles—to regulate susceptibility to oxidative damage and ferroptosis \citep{Szlasa2020}. Together, these are indications that membrane organization constitutes an additional fractal-like layer of transport heterogeneity. Scale-aware diffusion models that extend beyond homogeneous cytosolic or tissue-level assumptions is to be employed to explain the process.

Geometry also modulates radiobiological responses. The discovery of the \textit{FLASH} radiotherapy effect---characterized by remarkable normal-tissue sparing under ultra-high dose rate (UHDR $>40\,\mathrm{Gy/s}$) irradiation---has challenged traditional radiochemistry-based paradigms \citep{Vozenin2022,Pratx2019}. While oxygen depletion and radical recombination kinetics have been proposed as principal mechanisms \citep{MontayGruel2019,Labarbe2020,Cao2021,Boscolo2021,Favaudon2022,Espinosa2022}, emerging evidence indicates spatial topology and diffusion-channel connectivity are equally decisive. \citet{Abolfath2023} employed a stochastic reaction-diffusion model to simulate radiation-induced reactive oxygen species (ROS) transport in disordered geometries, showing that tissue-like media facilitate extensive inter-track recombination, leading to oxygen depletion and reduced damage (the FLASH-sparing effect). In contrast, fractal-like tumor architectures restrict ROS percolation, isolating reactive clusters and thereby preserving oxidative injury while abolishing sparing. Recently, \citet{Guo2024}  developed a uniform oxygen diffusion-reaction model that accounts for the oxygen depletion as a function of dose rate, arguing transient diffusion bottlenecks may accentuate tissue-type–specific dose-rate responses. In contrast, we incorporated geometry-governed localization effects that may bridge microscopic structural organization with macroscopic radiobiological outcomes to explain differential tissue response at UHDR. 
%Building on this, \citet{Guo2024} developed a microscopic oxygen diffusion-reaction model for UHDR, demonstrating transient diffusion bottlenecks that accentuate tissue-type–specific dose-rate responses. Geometry-governed localization effects thus bridge microscopic structural organization with macroscopic radiobiological outcomes.

In a nutshell, intertrack recombination of reactive species (RS) emerges if the particles constituting a therapeutic beam enter tissues within short spatial and temporal intervals ($< 0.1\mu m$, and $< 1 \mu s$). Under such conditions, the biological response of cells and tissues may exhibit sensitivity to dose rate, leading to lower reactivity rates to biomolecules such as lipids and DNA, simply because RS react with each other prior to reacting with biomolecules. This mutually inclusive condition can be broken down if the cellular blockage of the transport channels among the tracks prevents inter-track diffusion of RS. Thus, the reaction to tissue can only be described by means of an intra-track mechanisms, as the reactivity among RS and biomolecules would be dominant within independent and uncorrelated tracks. The time-lag among the tracks becomes an irrelevant parameter, and the biological response of cells and tissues shows insensitivity to radiation dose rate. Thus, the diffusibility of RS in tissues may strongly influence the differential tissue responses to radiation dose rate, i.e., FLASH-UHDR vs. conventional dose rate (CDR). 

Importantly, the transition from independent to interacting radiation tracks defines a regime in which transport properties become functionally relevant. Under CDRs, temporal separation between tracks limits the coexistence of RS clouds, and inter-track interactions are therefore suppressed. In contrast, under FLASH-UHDR irradiation, the high density of tracks exposes a regime in which simultaneous multi-track transport enables overlap, and geometry-governed diffusion becomes a controlling factor in radiochemical outcomes.

Collectively, these studies demonstrate that geometry and topology are fundamental determinants of therapeutic transport, as crucial as local biochemistry or kinetic processes. The diffusion field---the tissue itself---acts as an active modulator of oxygen, drug, and ionic/radical dynamics. Nevertheless, the prevailing radiobiological and pharmacokinetic models often assume homogeneous diffusivity and continuous concentration fields, limiting their predictive precision in realistic tissues \citep{Abolfath2023}.

To address this limitation, we develop a generalized diffusion framework grounded in fractal geometry, in which transport coefficients are intrinsically scale-dependent and reflect the underlying tissue morphology. A central feature of this framework is the introduction of a scaling exponent, $\theta$, that quantifies the degree of geometric complexity and anisotropy within the medium. This leads to a position-dependent RS diffusivity and/or ionic (electrical) conductivity
\begin{equation}
\sigma(r) \propto r^{-\theta},
\end{equation}
governing transitions from Euclidean to fractal transport regimes. The framework recovers classical Fickian diffusion in the limit $\theta \to 0$ where diffusivity and/or conductivity are uniformly constant (due to underlying translational symmetry), but yields anomalous (nonlinear) transport in structurally disordered domains where the RS diffusivity and/or electrical conductivity depend on the initial position of ions, with broken translational invariance. 

By defining an effective \textit{fractal conductivity}, %$\kappa_f$, 
we establish a quantitative link between microstructural disorder and macroscopic transport efficiency, allowing us to compare tissue architectures in terms of their diffusive impedance. Analytical solutions for spherical and cylindrical geometries—representative of tumor nodules and localized drug depots—illustrate how fractal geometry modulates oxygen delivery, drug penetration, and radical transport.

In the following sections, we formalize the fractal diffusion equations and derive their scaling relations, analyze representative solutions under biologically relevant geometries, and discuss implications for oxygen depletion and tissue sparing in \textit{FLASH} radiotherapy. This structure provides both a mechanistic foundation and a pathway toward geometry-aware predictive modeling of the therapeutic response.

\section{Methods}
To investigate the role of tissue geometry and structural disorder in molecular transport, we systematically compared three distinct diffusion descriptions:
(i) generalized fractal diffusion,
(ii) classical normal (Euclidean) diffusion, and
(iii) a Gaussian reference model.

All three models are evaluated under identical physical and geometrical conditions to isolate the effects of fractal dimensionality and anomalous transport from purely kinetic or chemical influences. This comparative framework allows us to elucidate how structural complexity, encoded through the fractal dimension \(D\) and the scaling exponent \(\theta\), modifies both transient and steady-state diffusion behavior relative to classical Fickian theory.

The formulation is intentionally minimal and geometry-driven, enabling direct interpretation of transport behavior in heterogeneous biological tissues without introducing additional phenomenological assumptions.

\subsection{Generalized fractal diffusion model}
Transport in heterogeneous biological tissues, embedded in Euclidean space with dimension $d$ is governed not only by local diffusivity but also by the geometry and connectivity of accessible pathways. In complex media such as tumors, diffusion pathways are tortuous, partially disconnected, and hierarchically organized, leading to deviations from classical Gaussian transport. To capture these effects, we model diffusion on a fractal substrate characterized by two independent indices of (1) a Hausdorff (fractal) dimension \(D\) and (2) a scaling exponent \(\theta\) that quantifies geometric resistance and anomalous transport.

The normalized radial probability density (distribution function), \(P(r,t)\), satisfies the generalized diffusion--reaction equation
\begin{equation}
\frac{\partial P(r,t)}{\partial t}
=
\frac{1}{r^{D-1}}
\frac{\partial}{\partial r}
\left[
k\, r^{D-1-\theta}
\frac{\partial P(r,t)}{\partial r}
\right]
-
\mu\, P(r,t),
\label{eq:fractal_diffusion_section2}
\end{equation}
where \(k\) is the homogeneous microscopic transport coefficient and \(\mu\) represents an effective reaction or decay rate.
The factor \(r^{D-1-\theta}\) introduces a scale-dependent reduction of diffusive flux, reflecting tortuosity and partial disconnection of transport pathways in fractal media.

At $\mu=0$, the probability density is normalized according to the fractal measure
\begin{eqnarray}
1 &=& \int d\hat{\Omega}\int_0^{\infty} P_{\hat{\Omega}}(r,t)\, r^{D-1} \, dr 
\nonumber \\ &=&
\int_0^{\infty} P(r,t)\, r^{D-1} \, dr,
\label{eq:fractal_norm}
\end{eqnarray}
which ensures conservation of probability within a domain of non-integer (fractional) dimensionality. 
$\hat{\Omega}$ is the solid angle.
In Eq. (\ref{eq:fractal_norm}), $P(r,t)$ is an angular independent PDF, hence the normalization factors stemming from the integration over $\hat{\Omega}$ have been absorbed in $P(r,t)$. 

In the limit \(\theta=0\), Eq.~(\ref{eq:fractal_diffusion_section2}) reduces to diffusion on a purely fractal geometry. For \(\theta>0\), additional geometric resistance and memory effects emerge, leading to anomalous (subdiffusive) transport. %This formulation provides a continuous description that bridges Euclidean and fractal transport regimes while retaining a clear physical interpretation in terms of tissue microarchitecture.

\subsection{Normal (Euclidean) diffusion}
Normal diffusion refers to Fickian transport characterized by a linear growth of the mean-squared displacement with time, \(\langle r^2(t) \rangle \propto t\).
It is recovered in the Euclidean limit with \(D = d\) and \(\theta=0\), for which Eq.~(\ref{eq:fractal_diffusion_section2}) reduces to
\begin{equation}
\frac{\partial P(r,t)}{\partial t}
=
k
\left[
\frac{1}{r^{d-1}}
\frac{\partial}{\partial r}
\left(
r^{d-1}
\frac{\partial P(r,t)}{\partial r}
\right)
\right]
-
\mu\, P(r,t).
\label{eq:normal_diffusion_section2}
\end{equation}
This equation describes diffusion in a homogeneous spherical or cylindrical geometry, corresponding to $d=3$ or 2, respectively. It serves as an appropriate reference model for transport in normal tissues, where extracellular spaces and intracellular environments are relatively uniform and well connected. In this regime, diffusion pathways are minimally constrained, and transport is well approximated by Fickian dynamics. The details of the numerical solution of the normal diffusion are given in Appendix \ref{app:numerics}.

\subsection{Gaussian reference model}
For comparison, we also include a Gaussian as a reference distribution. Note that Gaussian distribution is a solution of normal diffusion for a specific boundary condition such that at $t=0$, $P(r,t)=\delta(r)$ and at $r=\infty$ and finite $t$, $P(r,t)=0$. Denoting 
\begin{equation}
P_{\mathrm{G},d}(r,t)
=
\frac{1}{\mathcal{N}_d(t)}
\exp\!\left[-\frac{r^{2}}{4kt}\right],
\label{eq:gaussian_section2}
\end{equation}
where \(\mathcal{N}_d(t)=\frac{2}{\Gamma(d/2)} 
\left[ \frac{1}{4kt} \right]^{d/2}\) ensures normalization. Here $d$ is the Euclidean dimension.
$\Gamma(n)=(n-1)!$, 
%and $(2n - 1)!! = (2n - 1)(2n - 3) \cdots (3)(1)$
for integer $n$, and
$\Gamma \left(n+ \frac{1}{2}\right) =  \binom{n-\frac{1}{2}}{n} n! \sqrt{\pi}$.

\begin{figure}[t]
\begin{center}
\includegraphics[width=1.0\linewidth]{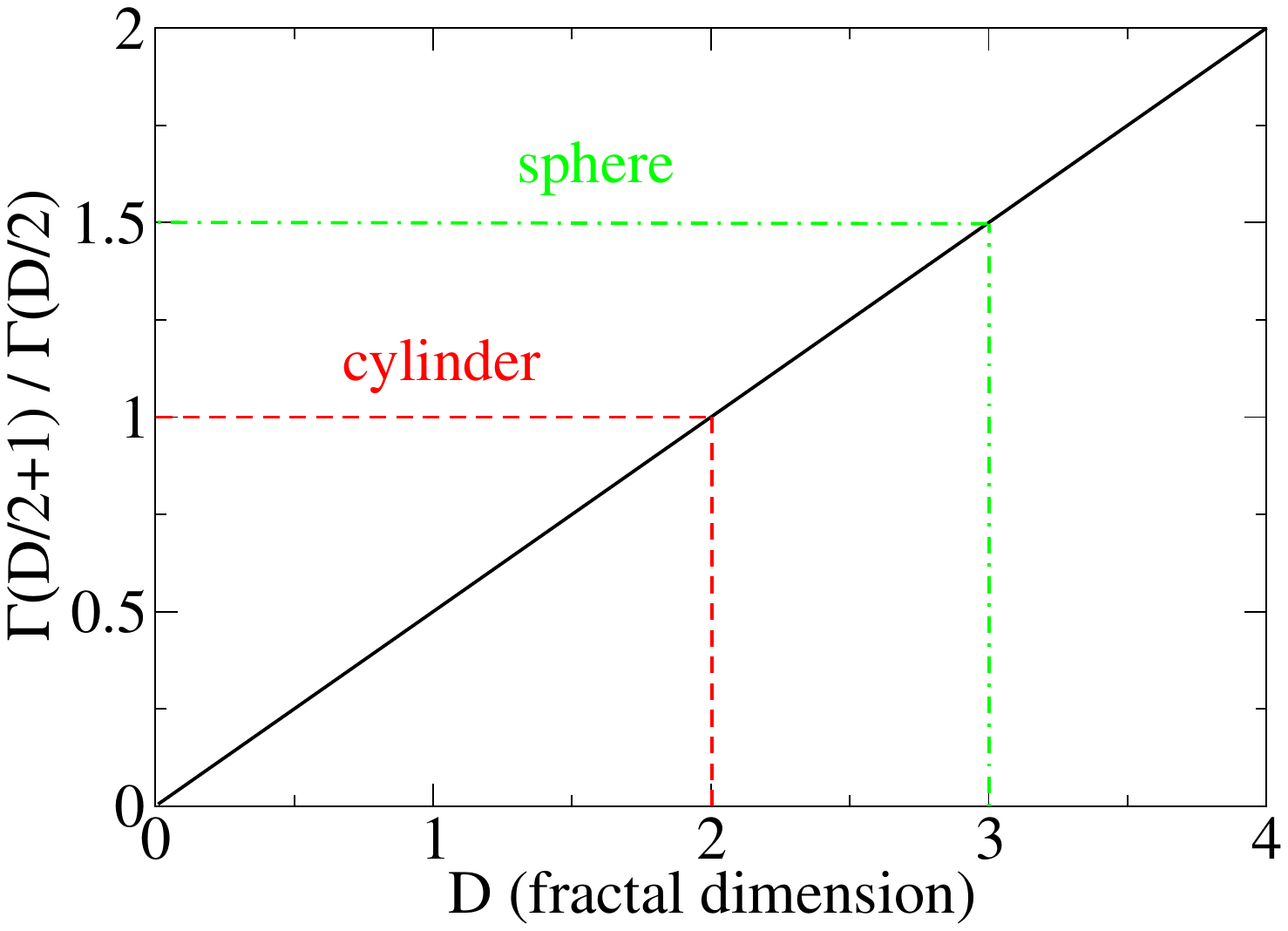}\\ %\vspace{-1.0cm}
\noindent
\caption{
    $\Gamma(D/2 +1)/\Gamma(D/2)$ vs. $D$. 
Two special limits of $D=2$ and $D=3$ correspond to cylindrical and spherical geometries.
}
\label{fig1}
\end{center}\vspace{-0.5cm}
\end{figure}

Considering this boundary condition, for $\theta > 0$ and $D\neq d$, we find a non-Gaussian distribution function as a solution of Eq. (\ref{eq:fractal_diffusion_section2})
\begin{align}
    P(r, t)=& \frac{2 + \theta}{\Gamma(D/(2 + \theta))} 
\left[ \frac{1}{k (2 + \theta)^2 t} \right]^{D/(2 + \theta)} \nonumber  \\ 
\times&\exp\left[-\frac{r^{2 + \theta}}{k (2 + \theta)^2 t} \right].
\label{eq14}
\end{align}
At $\theta=0$ the diffusion length follows the Einstein's relation, $\langle r^2\rangle \propto t$. However, for $D \neq d$, the normalization of $P_{G,D}(r,t) = \exp(-r^2/4kt)/\mathcal{N}_D(t)$ results in 
\(\mathcal{N}_D(t)=\frac{2}{\Gamma(D/2)} 
\left[ \frac{1}{4kt} \right]^{D/2}\) where
$\Gamma(z) = \int_{0}^{\infty} t^{z-1} e^{-t} \, dt$.
As the amplitude of $P_G(r, t)$ decreases with increasing $D$, one can find $P_{G,D} < P_{G,d}$ for $D > d$.  
From this PDF, it is straightforward to show
\begin{eqnarray}
\langle r^2(t) \rangle &=& \int_0^\infty dr r^{D-1} r^2 P(r, t) 
\nonumber \\ &=& 
\frac{\Gamma \left(\frac{D+2}{2+\theta}\right)}{\Gamma \left(\frac{D}{2+\theta}\right)} 
\left[k(2+\theta)^2 t\right]^{2/(2+\theta)}
\sim t^{2/(2+\theta)}. \nonumber  \\
\label{eq8}
\end{eqnarray}
The exponent \(\theta\) provides a quantitative and physically transparent measure of deviation from Fickian transport and serves as a key control parameter linking tissue microstructure to macroscopic transport behavior.
To quantify transport behavior across models, we analyze the mean-square displacement (MSD), which follows the scaling as in Eq. (\ref{eq8}).
In the limit of $\theta=0$, Eq.(\ref{eq8}) leads to $\langle r^2(t) \rangle = \Gamma(D/2 +1)/\Gamma(D/2) 4 k t$, thus $k \rightarrow \Gamma(D/2 +1)/\Gamma(D/2) k$, globally, with normal diffusion and the growth exponent, 1/2. 
Similarly, for $D=d$, Eq.(\ref{eq8}) yields correctly $2kt$ for each degree of freedom. 
See Fig. \ref{fig1}.
For $D=d=2$, $\langle r^2(t) \rangle = \langle x^2(t) \rangle + \langle y^2(t) \rangle$, and $\Gamma(D/2 +1)/\Gamma(D/2) = \Gamma(2)/\Gamma(1) = 1$ thus $\langle x^2(t) \rangle = \langle y^2(t) \rangle = 2 k t$, as expected. 
For $D=d=3$, $\Gamma(D/2 +1)/\Gamma(D/2) = \Gamma(2+1/2)/\Gamma(1+1/2) = (3 \sqrt{\pi}/2^2)/(\sqrt{\pi}/2) = 3/2$. As 
$\langle x^2(t) \rangle = \langle y^2(t) \rangle = \langle z^2(t) \rangle = \langle r^2(t) \rangle/3$ we find 
$\langle x^2(t) \rangle = (3/2)/3 \times 4 k t = 2 k t$, that is the correct uniform diffusion relation for each degree of freedom. 
With the decrease of $D$ from $d$, as shown in Fig. \ref{fig1}, $\Gamma(D/2 +1)/\Gamma(D/2)$ decreases. Note that $D < d$ corresponds to an increase in the medium porosity, where the global diffusion in the medium decreases, due to lower accessibility of the geometrical volume for the chemical transport.
Note that this model calculation does not capture a sharp drop in the diffusion constant at the percolation threshold, as shown in Ref. \cite{Abolfath2023}.    

The Gaussian PDF does not correspond to an exact solution of the diffusion equation for any other boundary conditions, such as the ones used in this work, e.g., a uniform distribution in a cylinder with radius $R_c$ at $t=0$. 
It has been included in our discussion to benchmark the numerical solutions and for illustrating the fastest possible spatial spreading in the absence of geometric constraints or anomalous effects. Deviations from Gaussian behavior therefore provide a direct and intuitive measure of the impact of structural disorder and anomalous transport, as in large distances ($r >> R_c$) and finite time, because the exact solutions of normal diffusion asymptotically approach the Gaussian.

\begin{figure*}[!htbp]
	\begin{subfigure}{0.32\textwidth}\includegraphics[width=\textwidth]{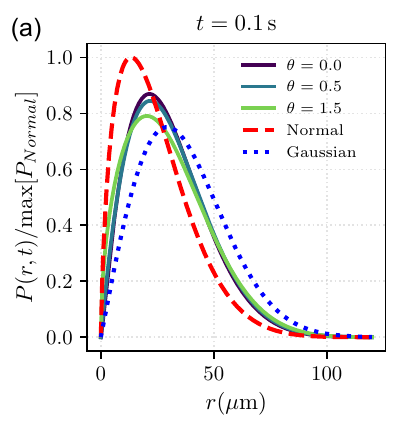}
	\end{subfigure}
	\begin{subfigure}{0.32\textwidth}\includegraphics[width=\textwidth]{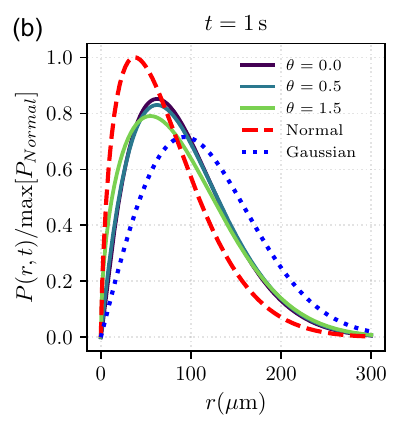}
	\end{subfigure}
	\begin{subfigure}{0.32\textwidth}\includegraphics[width=\textwidth]{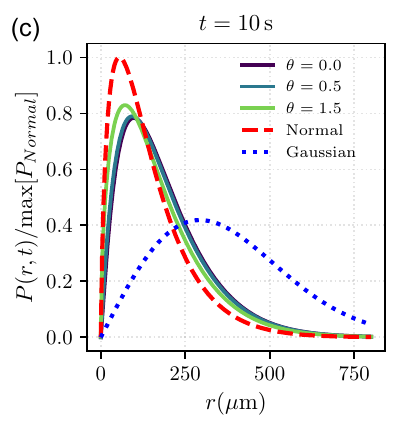}
	\end{subfigure}
	
	\caption{
Radial probability distribution \(P(r,t)\) as a function of distance \(r\) for the fractal diffusion model with fractal dimension \(D = 2.5\), shown for different values of the fractional parameter \(\theta = 0.0, 0.5,\) and \(1.5\).
Panels (a)--(c) correspond to times \(t = 0.1\,\mathrm{s}\), \(t = 1\,\mathrm{s}\), and \(t = 10\,\mathrm{s}\), respectively.
For comparison, the classical normal diffusion (red dashed line) and Gaussian diffusion (blue dotted line) solutions are also displayed.
Increasing \(\theta\) leads to enhanced localization and slower spreading relative to the normal and Gaussian cases, highlighting the influence of anomalous transport induced by fractal geometry.
}
	\label{fig2}
	%\end{center}
\end{figure*}

\begin{figure*}[!htbp]
	\begin{subfigure}{0.32\textwidth}\includegraphics[width=\textwidth]{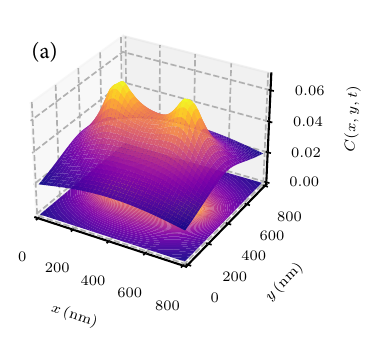}
	\end{subfigure}
	\begin{subfigure}{0.32\textwidth}\includegraphics[width=\textwidth]{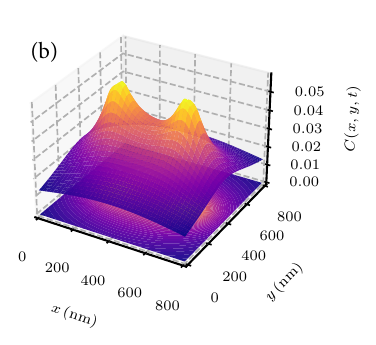}
	\end{subfigure}
	\begin{subfigure}{0.32\textwidth}\includegraphics[width=\textwidth]{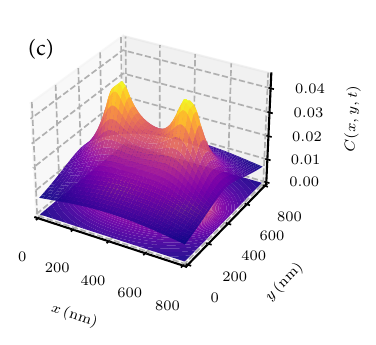}
	\end{subfigure}
	
	\caption{$D=2.5$ and $t=0.005s$ (a) $\theta=0.3$ (b) $\theta=0.5$ (c) $\theta=0.8$}

	\label{fig2-1}
	%\end{center}
\end{figure*}

\begin{figure*}[!htbp]
	\begin{subfigure}{0.23\textwidth}\includegraphics[width=\textwidth]{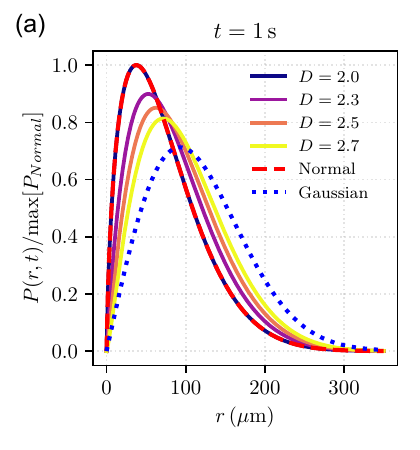}
	\end{subfigure}
    \begin{subfigure}{0.23\textwidth}\includegraphics[width=\textwidth]{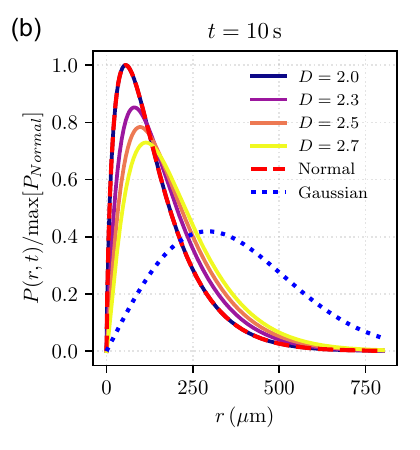}
	\end{subfigure}
	\begin{subfigure}{0.23\textwidth}\includegraphics[width=\textwidth]{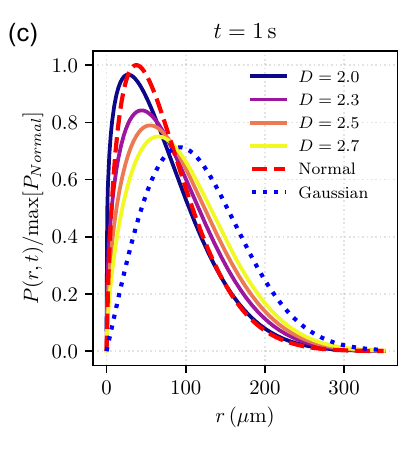}
	\end{subfigure}
	\begin{subfigure}{0.23\textwidth}\includegraphics[width=\textwidth]{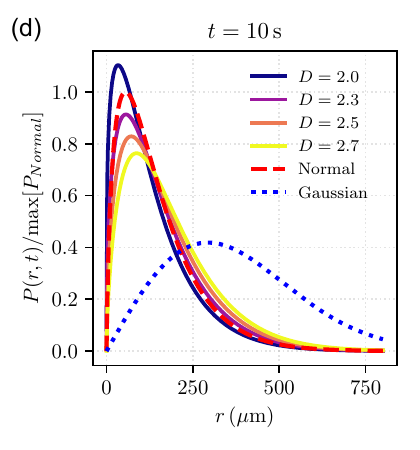}
	\end{subfigure}
	
	\caption{
Radial probability distribution \(P(r,t)\) as a function of distance \(r\) for the fractal diffusion model with varying fractal dimension \(D\).
Panels (a) and (b) correspond to the case \(\theta = 0\) at times \(t = 1\,\mathrm{s}\) and \(t = 10\,\mathrm{s}\), respectively, while panels (c) and (d) show the case \(\theta = 1.5\) at the same times.
Results are shown for different fractal dimensions \(D = 2.0, 2.3, 2.5,\) and \(2.7\).
For comparison, the classical normal diffusion solution with \(D = 2\) (red dashed line) and the Gaussian diffusion solution (blue dotted line) are also included.
The figure illustrates how both the fractal dimension and the fractional parameter \(\theta\) control the spreading rate and shape of the radial probability distribution.
}
	\label{fig3}
	%\end{center}
\end{figure*}

\begin{figure*}[!htbp]
	\begin{subfigure}{0.32\textwidth}\includegraphics[width=\textwidth]{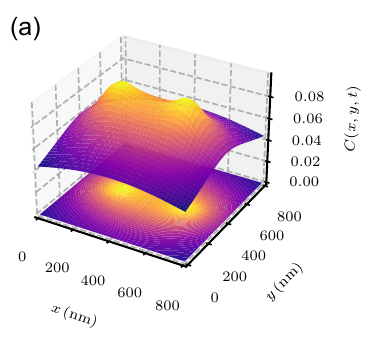}
	\end{subfigure}
	\begin{subfigure}{0.32\textwidth}\includegraphics[width=\textwidth]{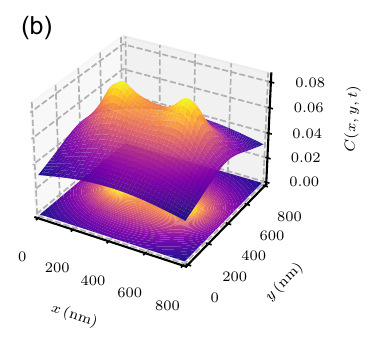}
	\end{subfigure}
	\begin{subfigure}{0.32\textwidth}\includegraphics[width=\textwidth]{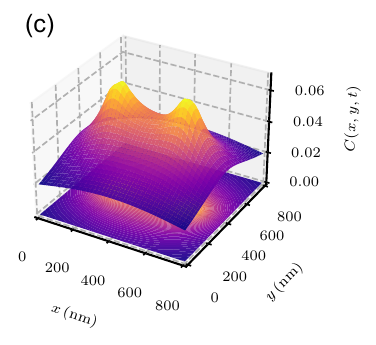}
	\end{subfigure}
	
	\caption{$\theta=0$ and $t=0.005s$ (a) $D=2$ (b) $D=2.5$ (c) $D=2.8$}

	\label{fig3-1}
	%\end{center}
\end{figure*}

\begin{figure*}[!htbp]
	%\begin{subfigure}{0.32\textwidth}\includegraphics[width=\textwidth]{P_rt_D_comparison_theta0.5_t1.0s.pdf}
	%\end{subfigure}
	\begin{subfigure}{0.32\textwidth}\includegraphics[width=\textwidth]{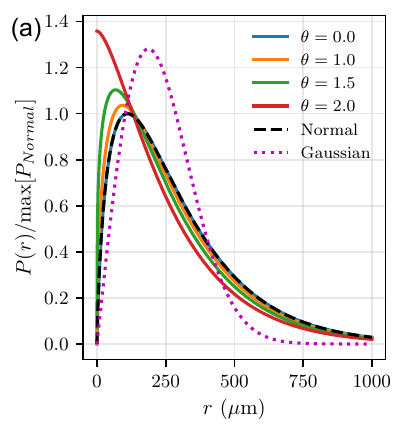}
	\end{subfigure}
	\begin{subfigure}{0.32\textwidth}\includegraphics[width=\textwidth]{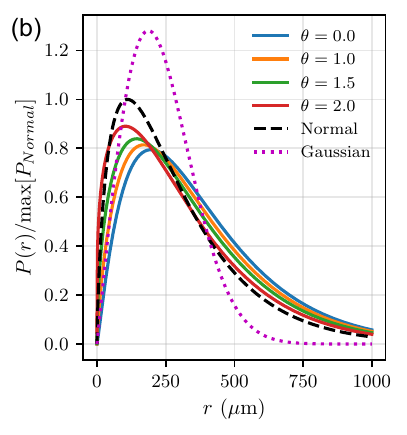}
	\end{subfigure}
	
	\caption{
Steady-state radial probability distribution \(P(r)\) as a function of distance \(r\) for the fractal diffusion model with different values of the fractional parameter \(\theta = 0.0, 1.0, 1.5,\) and \(2.0\).
Panel (a) corresponds to the Euclidean case with fractal dimension \(D = 2\), while panel (b) shows the fractal geometry with \(D = 2.5\).
For comparison, the steady-state solutions of normal diffusion (black dashed line) and Gaussian diffusion (magenta dotted line) are also presented.
The results demonstrate how both the fractal dimension and the parameter \(\theta\) modify the stationary spatial profile, leading to deviations from classical diffusive behavior.
}

	\label{fig4}
	%\end{center}
\end{figure*}

\begin{figure*}[!htbp]
	%\begin{subfigure}{0.32\textwidth}\includegraphics[width=\textwidth]{P_rt_D_comparison_theta0.5_t1.0s.pdf}
	%\end{subfigure}
	\begin{subfigure}{0.32\textwidth}\includegraphics[width=\textwidth]{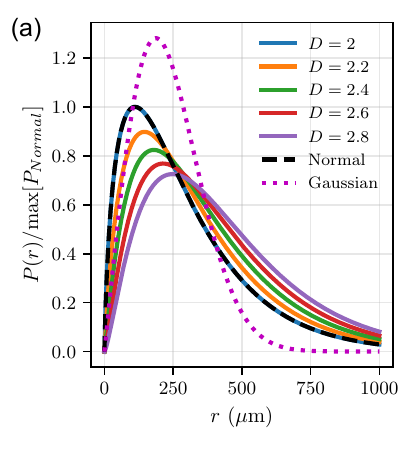}
	\end{subfigure}
	\begin{subfigure}{0.32\textwidth}\includegraphics[width=\textwidth]{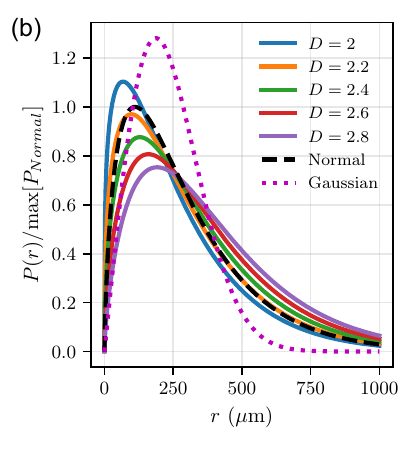}
	\end{subfigure}
	
	\caption{
Steady-state radial probability distribution \(P(r)\) as a function of distance \(r\) for the fractal diffusion model with varying fractal dimension \(D\).
Panel (a) corresponds to the case \(\theta = 0\), while panel (b) shows the case \(\theta = 1.5\).
Results are presented for different fractal dimensions \(D = 2.0, 2.2, 2.4, 2.6,\) and \(2.8\).
For comparison, the steady-state solutions of normal diffusion (black dashed line) and Gaussian diffusion 
(magenta dotted line) are also included.
The figure illustrates how the combined effects of the fractional parameter \(\theta\) and the fractal 
dimension \(D\) modify the stationary spatial distribution relative to classical diffusion models.
}

	\label{fig5}
	%\end{center}
\end{figure*}

To quantify inter-track interactions, we introduce an overlap measure between diffusive plumes generated by spatially separated sources. This quantity serves as a proxy for the probability of inter-track coupling, and therefore for the likelihood of radical recombination between tracks. By evaluating how this overlap depends on the parameters \(D\) and \(\theta\), we directly connect geometry-governed transport properties to inter-track interaction strength.

\subsection{Boundary conditions}
Throughout the entire work, we consider the numerical value of the diffusion constant $k=$1 nm$^2$/ns ($10^{-9}$ m$^2$/s). Note that for OH-radicals under thermal diffusion in water, $k=4.3$ nm$^2$/ns. 

All simulations shown in Figs.~\ref{fig2}–\ref{fig5} were performed under radially symmetric boundary conditions for Eq.~\ref{eq:fractal_diffusion_section2}. For the fractal and normal diffusion models, the computational domain was defined for $r \ge R_c$, where $R_c$ represents a microscopic cutoff length. At $r=R_c$, a mixed (Robin) boundary condition was imposed,
\begin{equation}
- k\,R_c^{D-1-\theta}
\left.\frac{\partial P(r,t)}{\partial r}\right|_{r=R_c}
=
\mathcal{P}\,[P_b(t)-P(R_c,t)],
\end{equation}
where $\mathcal{P}$ denotes an effective permeability parameter. The boundary value $P_b(t)$ was assumed constant in time, corresponding to $\hat P_b(s)=P_b/s$ in Laplace space ($s$ is the Laplace transform variable). 
At large distances, the domain was treated as unbounded with $P(r,t)\rightarrow0$ as $r\rightarrow\infty$, enforced through decaying modified Bessel-function solutions. In contrast, the Gaussian reference distribution corresponds to the fundamental solution of the normal diffusion equation in an infinite domain with initial condition $P(r,0)=\delta(r)$ and vanishing probability density at infinity, and therefore does not involve an explicit boundary condition at $r=R_c$. For steady-state calculations, identical spatial boundary conditions were applied with $\partial P/\partial t=0$, and all distributions were normalized using the appropriate geometrical measure.\\

%%%%%%%%%%%%%%%%%%%%%%%%%%%%%%%%%%%%%%%%%%%%%%%%%
\section{Results}
Table~\ref{tab:parameters} summarizes the fixed model parameters used in all numerical simulations, while control parameters such as the fractal exponent $\theta$ and $D$ are varied and reported separately in the corresponding figure captions.

\begin{table*}[!htbp]
\centering
\caption{Fixed model parameters used in all numerical simulations.}
\label{tab:parameters}
\begin{tabular}{l c c}
\hline\hline
Parameter & Description & Value \\
\hline\\
$k$ (or $\sigma$) & Diffusion coefficient (Conductivity)& $4.3\times10^{-9}\,\mathrm{m^2\,s^{-1}}$ \\
$k$ (or $\sigma$) (steady-state) & Diffusion coefficient (Conductivity)& $17.5\times10^{-9}\,\mathrm{m^2\,s^{-1}}$ \\
$\mu$ & Effective reaction (decay) rate & $0.5 \,\mathrm{s^{-1}}$ \\
$R_c$ & Microscopic cutoff radius & $0.5\times10^{-9}\,\mathrm{m}$ \\
$\mathcal{P}$ & Boundary permeability parameter & $1.5\times10^{-6}\,\mathrm{m}$ \\
%$D$ & Fractal (Hausdorff) dimension & $2.5$ \\
%$t$ & Observation time (transient profiles) & $5\,\mathrm{s}$ \\
%$P_b$ & Boundary probability density & $0.4$ \\
\hline\hline
\end{tabular}
\end{table*}

\subsection{Comparison of transient diffusion profiles}
Figures~\ref{fig2}–\ref{fig3} present the temporal evolution of the radial probability density \(P(r,t)\) for 
the three diffusion models—fractal, normal, and Gaussian—under identical conditions. Across all models, the 
distributions are initially localized near the origin at early times, reflecting the common initial 
condition. As time progresses, however, systematic differences in both spreading rate and profile shape 
emerge.

The influence of the fractional parameter \(\theta\) is isolated in Figure~\ref{fig2} for a fixed fractal
dimension \(D=2.5\). 
At early times, variations in \(\theta\) primarily modify the peak height of the distribution, while the 
overall spatial extent remains comparable across cases. With increasing time, the effect of \(\theta\) 
becomes progressively more pronounced: larger values of \(\theta\) lead to enhanced localization, steeper 
decay of the distribution tail, and a marked suppression of long-distance transport. Relative to the normal 
and Gaussian solutions, increasing \(\theta\) produces heavier central accumulation and shorter effective 
diffusion lengths, indicating a systematic reduction in transport efficiency.\\

\begin{figure}[htbp]
    \centering
    \includegraphics[width=0.45\textwidth]%{Overlap_integral_d100nm_t10s.pdf}
    {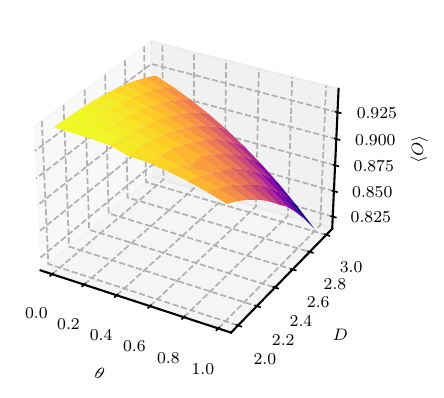}
    \caption{Three-dimensional visualization of the average overlap integral $O(t)$ as a function of the fractal dimension \(D\) and the anomalous diffusion exponent \(\theta\), evaluated at a fixed time \(t = 10\,\mathrm{s}\) and track separation \(400\,\mathrm{nm}\). }
    \label{fig:overlap_integral}
\end{figure}

Let $C(\mathbf{x},t;\mathbf{r}_i)$ denote the concentration field evaluated at observation point $\mathbf{x}$ and time $t$, generated by a point source located at position $\mathbf{r}_i$. We consider two identical sources placed at $\mathbf{r}_1$ and $\mathbf{r}_2$ within the same fractal medium characterized by transport parameters $D$ and $\theta$. The corresponding concentration profiles are therefore $C_1(\mathbf{x},t)=C(\mathbf{x},t;\mathbf{r}_1)$ and $C_2(\mathbf{x},t)=C(\mathbf{x},t;\mathbf{r}_2)$. Since both fields satisfy the same generalized diffusion equation under identical boundary conditions, they differ only by a spatial translation associated with the source positions. Defining the separation vector $\boldsymbol{\ell}=\mathbf{r}_2-\mathbf{r}_1$, translational invariance implies $C_2(\mathbf{x},t)=C_1(\mathbf{x}-\boldsymbol{\ell},t)$. The overlap integral is then defined as
\[
O(t)=\int_{V} C_1(\mathbf{x},t)\,C_2(\mathbf{x},t)\, d^2\mathbf{x},
\]
where $\mathbf{x}$ is the integration (observation) coordinate over the spatial domain $V$. This quantity measures the spatial coexistence of two otherwise identical diffusion profiles displaced by $\boldsymbol{\ell}$ and provides a direct macroscopic indicator of inter-source (inter-track) interaction and serves as a proxy for the probability of inter-track coupling.

Figure~\ref{fig2-1} shows the overlap integral $O(t)$ as a function of the anomalous exponent $\theta$, for a fixed fractal dimension $D = 2.5$ and time $t = 10\,\mathrm{s}$. The figure focuses on the integrated overlap, which quantifies the total spatial superposition of the two diffusive plumes.

For small $\theta$ [Fig.~\ref{fig2-1}(a)], the overlap integral attains relatively high values, indicating extended plume interaction consistent with weakly anomalous diffusion. As $\theta$ increases to an intermediate value [Fig.~\ref{fig2-1}(b)], $O(t)$ decreases noticeably, reflecting enhanced confinement and reduced plume overlap. In the strongly anomalous regime [Fig.~\ref{fig2-1}(c)], the overlap integral drops sharply, signifying severe suppression of inter-plume coupling due to highly localized transport.

These results confirm that increasing $\theta$ not only shortens the characteristic diffusion length but also systematically reduces the overall spatial overlap between plumes. Importantly, this effect is non-perturbative: even at a fixed fractal dimension, the transport efficiency—and consequently the overlap integral—can be dramatically reduced solely through the anomalous scaling exponent $\theta$. This highlights the dominant role of fractional dynamics in controlling mass transfer in structurally complex media.

Figure~\ref{fig3} examines the combined influence of the fractal dimension \(D\) and the fractional parameter
\(\theta\). For \(\theta=0\) [Figs.~\ref{fig3}(a,b)], increasing \(D\) leads to progressively broader 
distributions, reflecting enhanced spatial accessibility and convergence toward the classical normal 
diffusion limit as \(D \to 2\). In contrast, for \(\theta=1.5\) [Figs.~\ref{fig3}(c,d)], this trend is 
significantly attenuated: even at larger values of \(D\), the distributions remain comparatively localized. 
These results demonstrate that fractional dynamics can dominate over geometric connectivity in determining 
transient transport behavior, giving rise to a broad range of diffusion profiles not captured by standard 
diffusion models.

Figure \ref{fig3-1} presents two-dimensional steady-state spatial distributions for the case $\theta = 0$ at fixed time $t = 10\,\mathrm{s}$ while varying the fractal dimension $D$. In contrast to Figure 2, where the fractional parameter was varied at constant $D$, here the influence of geometric topology alone is isolated.

For the Euclidean case $D = 2$ [Fig. \ref{fig3-1}(a)], the distribution exhibits the broadest spatial spreading, consistent with classical diffusion behavior. As the fractal dimension increases to $D = 2.5$ and $D = 2.8$ [Figs. \ref{fig3-1}(b,c)], the spatial profile progressively narrows and peak localization increases. This behavior reflects the reduction of effective accessible volume despite the nominal increase in geometric dimension, a characteristic feature of fractal substrates where connectivity and tortuosity compete with dimensional scaling.

Importantly, even in the absence of fractional effects ($\theta = 0$), purely geometric fractality modifies the spatial organization of the probability density. This demonstrates that topology alone can alter transport characteristics, independent of anomalous temporal scaling. When combined with nonzero $\theta$ (as shown in earlier figures), these geometric effects become even more pronounced, reinforcing the complementary roles of $D$ and $\theta$ in controlling diffusion dynamics.

\subsection{Steady-state distributions}
The steady-state radial probability distributions are shown in Figure~\ref{fig4} for varying values of
\(\theta\) and in Figure~\ref{fig5} for varying values of \(D\). In Figure~\ref{fig4}(a), corresponding to 
the Euclidean case \(D=2\), the steady-state distribution closely follows the normal diffusion solution for 
small values of \(\theta\), with noticeable deviations emerging only at larger \(\theta\). As \(\theta\) 
increases, the distribution exhibits reduced peak height and increasingly pronounced tails, indicating a 
departure from classical equilibrium behavior.

In contrast, Figure~\ref{fig4}(b), corresponding to the fractal geometry with \(D=2.5\), shows that even for 
moderate values of \(\theta\), the steady-state distributions differ substantially from both normal and 
Gaussian solutions. The presence of broader tails and suppressed peaks indicates that fractal geometry alone 
is sufficient to alter the stationary spatial organization. Comparison with the Gaussian and normal steady-
state solutions confirms that the equilibrium distributions retain a clear dependence on the underlying 
transport mechanism.

Figure~\ref{fig5} further elucidates the role of the fractal dimension \(D\) in the stationary regime. 
For \(\theta=0\), Figure~\ref{fig5}(a), increasing \(D\) produces progressively wider distributions, 
consistent with increased spatial connectivity. For \(\theta=1.5\), Figure~\ref{fig5}(b), the same increase 
in \(D\) results in significantly weaker broadening, demonstrating that fractional dynamics dominate over 
geometric effects in determining the steady-state profile. Together, these results indicate that \(D\) and
\(\theta\) play complementary but distinct roles: \(D\) governs the spatial topology of the medium, while
\(\theta\) controls the effective transport efficiency across scales.\\

\section{Discussion}

The results presented in this study demonstrate that diffusion in structurally heterogeneous media is 
fundamentally governed by geometry, and that deviations from classical Fickian behavior emerge naturally when
transport occurs on fractal substrates. While all 
diffusion models considered here exhibit strong localization at early times due to identical initial 
conditions, pronounced differences arise as transport evolves. The Gaussian reference model exhibits the 
fastest spatial spreading, followed by classical normal diffusion, whereas the generalized fractal diffusion
model consistently shows slower radial expansion and enhanced localization. This hierarchy of spreading rates 
reflects increasing geometric constraints and reduced connectivity of transport pathways.

At the mechanistic level, this behavior originates from the anomalous flux term
\(r^{D-1-\theta}\partial_r P\), which introduces a scale-dependent suppression of transport at larger 
distances. As a result, long-range diffusion becomes increasingly inefficient, leading to subdiffusive 
dynamics characterized by heavy central accumulation and truncated spatial tails. These features persist 
across both transient and steady-state regimes, indicating that anomalous transport is an intrinsic 
consequence of geometry rather than a transient kinetic effect \cite{Metzler2000,Zaslavsky2002}.

A pronounced sharp peak appears in the spatial concentration field in the immediate vicinity of the track core, originating from the small-argument behavior of the modified Bessel functions entering the Green-function kernel. In particular, when $qr \to 0$ (which occurs at very small radial distances or at early times), the modified Bessel function of the second kind, $K_\nu(qr)$, exhibits strong amplification that reflects the singular character of an idealized point-like source. Although this behavior is mathematically consistent with the analytical structure of the solution, it produces highly localized spikes in the numerical implementation that are not physically meaningful at the resolved scale of the model and may lead to numerical instability. To regularize this short-distance singularity, we introduced a smooth cutoff by replacing the argument $qr$ with $\sqrt{(qr)^2 + \varepsilon^2}$. This modification prevents the kernel from probing arbitrarily small spatial scales, effectively rounding the Bessel-driven divergence while preserving the correct asymptotic behavior at larger distances. Physically, the cutoff can be interpreted as accounting for the finite core radius of the track or the limited spatial resolution of the medium, thereby removing the unphysical divergence associated with a strictly point-like source. Importantly, this regularization suppresses the artificial peak without significantly altering the large-scale spatial distribution or the integrated overlap, which remains governed by contributions from physically relevant distances.

A key outcome of this work is the clear separation of roles played by the fractal dimension \(D\) and the 
fractional parameter \(\theta\). The fractal dimension encodes the spatial topology and connectivity of the 
medium, controlling the degree to which space is accessible for diffusion 
\cite{Weibel1962,Baish2000,Gazit1995}. In contrast, \(\theta\) governs temporal transport efficiency, 
capturing memory effects, transient trapping, crowding, and heterogeneous transport resistance along diffusion 
pathways. The competition between these two parameters gives rise to a wide spectrum of diffusion behaviors, 
ranging from near-Fickian transport to strongly localized, non-Gaussian dynamics.

This separation provides a physically transparent framework for interpreting tissue heterogeneity. Media
characterized by near-Euclidean geometry and low \(\theta\) values support efficient transport and rapid 
homogenization, leading to diffusion profiles close to classical normal or Gaussian limits. Conversely, media 
with elevated fractal dimensions and nonzero \(\theta\) exhibit suppressed long-range transport, enhanced 
localization, and long-lived concentration gradients. Importantly, these effects persist in the steady state,
demonstrating that equilibrium distributions retain a memory of the underlying geometric disorder, as 
previously suggested in studies of tumor vasculature and porous biological networks 
\cite{Gazit1997,Guiot2006}.

When viewed in a biological context, these transport regimes naturally map onto differences between normal 
and tumor tissues. Normal tissues are typically characterized by relatively homogeneous extracellular spaces,
well-connected intracellular environments, and weak geometric constraints, conditions under which transport 
is often well approximated by classical or weakly anomalous diffusion \cite{Weibel1962,Grimes2014}. In 
contrast, tumor tissues exhibit pronounced architectural irregularity arising from abnormal cell packing, 
chaotic vasculature, dense extracellular matrix remodeling, and elevated tortuosity 
\cite{Baish2000,Gazit1997,Vaupel2004}. These features reduce the accessible transport pathways and introduce 
scale-dependent resistance, conditions that are naturally captured by a fractal description with \(D>2\) and 
nonzero \(\theta\), consistent with experimentally observed diffusion-limited hypoxia and heterogeneous drug 
penetration \cite{Vaupel2021,Degner1988,Stylianopoulos2013}.

The relevance of this framework is further reinforced by experimental observations of membrane-scale 
heterogeneity in cancer cells. Fluorescence lifetime imaging and viscosity-sensitive probes have revealed 
highly heterogeneous nanoviscosity landscapes within lipid bilayers, even within a single membrane 
\cite{Ober2019}. Tumor cell membranes, in particular, exhibit significantly higher and more spatially 
variable viscosities compared to normal cells \cite{Li2023}. Such nanoscale mechanical heterogeneity 
introduces additional resistance to lateral diffusion and reaction--diffusion coupling, promoting non-
Gaussian transport behavior. Within the present framework, these effects are naturally captured by nonzero
values of \(\theta\) and elevated effective fractal dimensions, linking membrane-scale disorder to 
macroscopic transport anomalies \cite{Szlasa2020}.

From a radiobiological perspective, these findings have direct implications for understanding tissue-
dependent responses to ultra-high dose rate (FLASH) irradiation. The localization of reactive species,
oxygen, and signaling molecules within fractal-like media implies reduced spatial overlap and diminished 
inter-track interactions, particularly in structurally complex tumor tissues \cite{Abolfath2023,Guo2024}, consistent with the overlap suppression quantified in the present model. In
contrast, in more homogeneous media, enhanced connectivity promotes inter-track overlap, facilitating 
recombination processes and rapid spatial homogenization, a mechanism consistent with recent stochastic 
reaction--diffusion and track-structure analyses of FLASH response \cite{Boscolo2021,Abolfath2023}.

The present results indicate that geometry does not create inter-track interactions, but rather controls their extent once multiple tracks coexist within short spatial and temporal separations. In this sense, FLASH irradiation can be interpreted as a regime in which simultaneous multi-track transport makes overlap a relevant quantity, thereby exposing the role of geometry in governing reactive-species coupling. Under conventional dose-rate conditions, where such coexistence is limited, the same transport properties persist but have a reduced impact on inter-track interactions. Importantly, this framework does not replace existing biochemical or radiochemical explanations of FLASH effects, such as oxygen depletion and radical recombination kinetics \cite{Pratx2019,Labarbe2020,Favaudon2022}, but rather complements them by identifying tissue geometry as an active modulator of transport and reactivity. By explicitly incorporating scale-dependent diffusivity and fractal topology, the model provides a unifying description that bridges microscopic structural disorder with macroscopic transport efficiency. This geometry-driven perspective helps reconcile disparate experimental observations, including hypoxia, delayed drug penetration, heterogeneous oxidative stress, and reduced inter-track chemical recombination in malignant tissues \cite{MontayGruel2019,Vozenin2022,Espinosa2022}.

Overall, the generalized fractal diffusion model establishes a mechanistic link between tissue architecture 
and transport behavior, demonstrating that anomalous diffusion is not merely a mathematical abstraction but a 
physically necessary ingredient for describing transport in heterogeneous biological systems. By framing 
diffusion as a geometry-governed process, this work provides a foundation for geometry-aware modeling of 
radiobiological response and motivates future experimental efforts to quantify tissue fractality, transport 
heterogeneity, and scale-dependent diffusivity in the context of FLASH radiotherapy.

\section{Conclusion}
In this work, we developed a geometry-governed diffusion framework to describe molecular transport in 
structurally heterogeneous biological tissues, with particular relevance to ultra-high dose rate (FLASH) 
radiotherapy. By formulating diffusion on fractal substrates and introducing a scaling exponent that captures 
geometric resistance and memory effects, we demonstrated how deviations from classical Fickian transport 
arise naturally from tissue architecture itself.

The model identifies two complementary parameters that control transport behavior: the fractal dimension
\(D\), which encodes the spatial topology and connectivity of the medium, and the fractional parameter
\(\theta\), which governs scale-dependent transport efficiency and anomalous dynamics. Together, these 
parameters define distinct diffusion regimes that persist across both transient and steady-state conditions.
Media characterized by near-Euclidean geometry and weak anomalous effects exhibit efficient spreading and 
rapid homogenization, while fractal-like media display suppressed long-range transport, enhanced 
localization, and non-Gaussian spatial profiles.

Within this framework, normal and tumor tissues naturally map onto different transport regimes. Normal 
tissues are associated with lower effective fractal dimensionality and small \(\theta\), leading to diffusion
behavior close to classical or weakly anomalous limits. In contrast, tumor tissues—characterized by architectural 
disorder, irregular vasculature, dense extracellular matrix remodeling, and elevated crowding—are well 
described by larger effective fractal dimensions and nonzero \(\theta\), resulting in localized diffusion and
long-lived concentration gradients. These transport characteristics provide a geometric explanation for 
experimentally observed phenomena such as chronic hypoxia, delayed drug penetration, and reduced spatial 
overlap of reactive species in malignant tissues.

From a radiobiological perspective, the present results provide a geometry-driven framework for interpreting differential 
tissue response under FLASH irradiation. By limiting inter-track overlap, as quantified by the overlap integral, and suppressing long-range transport
of radiolytic species in structurally complex media, fractal tissue geometry reduces collective chemical 
reactivity, while more homogeneous tissues permit greater inter-track interaction and recombination. 
Importantly, this mechanism arises from geometry-governed transport properties and does not rely on specific assumptions about 
biochemical reaction rates or oxygen depletion kinetics.

Overall, this study establishes fractal geometry as a fundamental determinant of transport efficiency in 
biological tissues and provides a unifying theoretical framework linking microscopic structural disorder to 
macroscopic radiobiological outcomes. By framing diffusion as a geometry-controlled process, the model 
complements existing chemical and kinetic descriptions of FLASH effects and motivates future experimental
efforts to quantify tissue fractality, transport heterogeneity, and scale-dependent diffusivity in both 
normal and malignant tissues. Such geometry-aware modeling may ultimately contribute to more predictive and 
tissue-specific approaches in radiotherapy planning and optimization.

\section{Acknowledgment}
RA was supported by the American Cancer Society Diversity in Cancer Research Institutional Development Grant (DICRIDG-21-074-01-DICRIDG) at Howard University.

\appendix
\section{Numerical Solution of the Diffusion Models}
\label{app:numerics}

This Appendix provides a detailed description of the numerical procedures used to obtain the transient radial probability distributions for the three diffusion models considered in this work: generalized fractal diffusion, normal (Euclidean) diffusion, and a Gaussian reference solution. All notation and normalization conventions are fully consistent with those introduced in Secs.~II and III.

%-------------------------------------------------
\subsection{Generalized Fractal Diffusion Model}

The generalized diffusion equation on a fractal substrate is given by Eq.~(3) of the main text,
\begin{equation}
\frac{\partial P(r,t)}{\partial t}
=
\frac{1}{r^{D-1}}
\frac{\partial}{\partial r}
\left[
k\, r^{D-1-\theta}
\frac{\partial P(r,t)}{\partial r}
\right]
- \mu P(r,t),
\label{eq:app_fractal_pde}
\end{equation}
where $D$ is the fractal (Hausdorff) dimension, $\theta$ is the fractional transport exponent, $k$ is the effective transport coefficient, and $\mu$ is an effective decay or reaction rate.

%-------------------------------------------------
\subsubsection{Laplace-space formulation}

To facilitate numerical evaluation, Eq.~(\ref{eq:app_fractal_pde}) is transformed to Laplace space,
\begin{equation}
\hat{P}(r,s)=\int_{0}^{\infty} e^{-st} P(r,t)\,dt,
\end{equation}
which yields
\begin{equation}
\left[
\frac{d^{2}}{dr^{2}}
+
\frac{D-1-\theta}{r}\frac{d}{dr}
-
q^{2}
\right]\hat{P}(r,s)
=
-\frac{P(r,0)}{k\, r^{D-1}},
\end{equation}
with
\begin{equation}
q=\sqrt{\frac{s+\mu}{k}},
\qquad
\nu=\frac{D-2+\theta}{2}.
\end{equation}

The corresponding Green’s function is
\begin{equation}
G(r,r';s)
=
r'^{-D+2+\theta}
\begin{cases}
I_{\nu}(qr)\,K_{\nu}(qr'), & r \le r',\\[4pt]
I_{\nu}(qr')\,K_{\nu}(qr), & r > r',
\end{cases}
\label{eq:app_green}
\end{equation}
where $I_{\nu}$ and $K_{\nu}$ denote modified Bessel functions of the first and second kind.

%-------------------------------------------------
\subsubsection{Particular solution and boundary condition}

The Laplace-space solution is written as
\begin{equation}
\hat{P}(r,s)
=
\hat{P}_{\mathrm{part}}(r,s)
+
A(s)\,K_{\nu}(qr),
\end{equation}
with the particular solution
\begin{equation}
\hat{P}_{\mathrm{part}}(r,s)
=
-\int_{R_c}^{\infty}
G(r,r';s)
\frac{P(r',0)}{k}\,
r'^{D-1}\,dr'.
\end{equation}

At the inner cutoff radius $r=R_c$, a Robin-type boundary condition is imposed,
\begin{equation}
- k\, R_c^{D-1-\theta}
\left.
\frac{\partial \hat{P}}{\partial r}
\right|_{r=R_c}
=
\Lambda
\left[
\hat{P}_b(s)-\hat{P}(R_c,s)
\right],
\label{eq:app_robin}
\end{equation}
where $\Lambda$ is an effective permeability parameter and $\hat{P}_b(s)$ is the Laplace-transformed boundary value. Equation~(\ref{eq:app_robin}) uniquely determines the amplitude $A(s)$.

%-------------------------------------------------
\subsubsection{Inverse Laplace transform}

The time-dependent probability density is obtained via numerical inversion of the Laplace transform using the Stehfest algorithm\cite{stehfest1970},
\begin{equation}
P(r,t)
=
\frac{\ln 2}{t}
\sum_{k=1}^{N}
V_k\,
\hat{P}\!\left(r, s_k\right),
\qquad
s_k=\frac{k\ln 2}{t},
\end{equation}
where $V_k$ are the Stehfest weights. Unless otherwise stated, $N=10$ is used, and convergence was verified by varying $N$.

Normalization is enforced according to the fractal measure,
\begin{equation}
\int_{R_c}^{\infty}
P(r,t)\, r^{D-1}\,dr = 1.
\end{equation}

%-------------------------------------------------
\subsection{Normal (Euclidean) Diffusion}

The normal diffusion model is recovered in the Euclidean limit $D=2$ and $\theta=0$. Equation~(\ref{eq:app_fractal_pde}) reduces to
\begin{equation}
\frac{\partial P(r,t)}{\partial t}
=
k
\left[
\frac{1}{r}
\frac{\partial}{\partial r}
\left(
r\frac{\partial P}{\partial r}
\right)
\right]
-
\mu P(r,t).
\end{equation}

In Laplace space, the Green’s function takes the form
\begin{equation}
G(r,r';s)
=
I_0\!\left(q\min\{r,r'\}\right)
K_0\!\left(q\max\{r,r'\}\right),
\end{equation}
with $q=\sqrt{(s+\mu)/k}$. The solution procedure, boundary condition, and Stehfest inversion are identical to those used for the fractal model.

Normalization is performed using the Euclidean measure,
\begin{equation}
\int_{0}^{\infty}
2\pi r\, P(r,t)\,dr = 1.
\end{equation}

%-------------------------------------------------
\subsection{Gaussian Reference Solution}

For comparison, we also consider the analytical Gaussian solution describing free diffusion in two dimensions,
\begin{equation}
P_{\mathrm{G}}(r,t)
=
\frac{1}{4\pi k t}
\exp\!\left(-\frac{r^{2}}{4kt}\right),
\end{equation}
which is normalized analytically. This solution represents the limiting case of homogeneous, memoryless transport and provides a reference for assessing deviations induced by fractal geometry and anomalous scaling.\\

%-------------------------------------------------

All three models are evaluated using identical transport parameters and initial conditions. Radial probability distributions are compared at fixed times to isolate the effects of fractal dimensionality $D$ and anomalous scaling $\theta$. Numerical quadrature is performed with adaptive integration, and normalization is verified to within numerical precision for all cases.

%--------------------------------------------------

\newpage
%\bibliography{refs}

%\bibliography{MyReferences}

 \end{document}